\documentclass{aastex}
\usepackage{emulateapj5}
\usepackage{apjfonts}

\def\kms{{\rm km}\,{\rm s}^{-1}} 

\newcommand{\ha}{${\rm H}\alpha\,$}
\newcommand{\be}{\begin{equation}}
\newcommand{\ee}{\end{equation}}
\newcommand{\masyr}{{\rm mas}\,{\rm yr}^{-1}}
\newcommand{\vrad}{v_{\rm rad}}
\newcommand{\ppc}{{\rm pc}^{-3}}
\newcommand{\mpc}{{M_{\odot}}{\rm pc}^{-3}}
\newcommand{\vmax}{{\mathbf V}_{\rm max}}

\begin{document}

\submitted{Accepted to ApJ}

\title{Cool White Dwarfs Revisited -- New Spectroscopy and Photometry}

\author{Samir Salim\altaffilmark{1,2}, R.\ Michael Rich\altaffilmark{1,2},
Brad M.\ Hansen\altaffilmark{1}, L.\ V.\ E.\ Koopmans\altaffilmark{3},
Ben R.\ Oppenheimer\altaffilmark{4} and Roger D.\ Blandford\altaffilmark{5}}

\altaffiltext{1}{Department of Physics and Astronomy, University of
California at Los Angeles, Los Angeles, CA 90095, samir@astro.ucla.edu, 
rmr@astro.ucla.edu, hansen@astro.ucla.edu}
\altaffiltext{2}{Visiting Astronomer, Lick Observatory}
\altaffiltext{3}{Space Telescope Science Institute, Baltimore, MD 21218, 
koopmans@stsci.edu}
\altaffiltext{4}{Astrophysics Department, American Museum of Natural History, 
New York, NY 10024, bro@amnh.org}
\altaffiltext{5}{Theoretical Astrophysics, California Institute of Technology,
Pasadena, CA 91125, rdb@tapir.caltech.edu}

\begin{abstract}
In this paper we present new and improved data on 38 cool white dwarfs
identified by \citet{ohdhs} (OHDHS) as candidate dark halo objects.
Using the high-resolution spectra obtained with LRIS on Keck I, we
measure precise radial velocities for 13 white dwarfs that show an \ha
absorption line. We show that the knowledge of radial velocities on
average decreases the $UV$-plane velocities by only 6\%. In two cases
the radial velocities put original halo candidates below the OHDHS
velocity cut. The radial velocity sample has a velocity dispersion in
the direction perpendicular to the Galactic plane of $\sigma_W =
59\,\kms$ -- in between the values typically associated with the thick
disk and the stellar halo populations. We also see indications for the
presence of two populations by analyzing the velocities in the $UV$
plane. In addition, we present CCD photometry for half of the sample,
and with it recalibrate the photographic photometry of the remaining
white dwarfs. Using the new photometry in standard bands, and by
applying the appropriate color-magnitude relations for hydrogen and
helium atmospheres, we obtain new distance estimates. By recalibrating
the distances of the white dwarfs that were not originally selected as
halo candidates, we obtain 13 new candidates (and lose 2 original
ones). On average, new distances produce velocities in the $UV$ plane
that are larger by 10\%, with already fast objects gaining more.
Using the new data, while applying the same $UV$-velocity cut
($94\,\kms$) and methods of analysis as in OHDHS, we find a density of
cool white dwarfs of $1.7\times 10^{-4}\,\ppc$, confirming the value
of OHDHS. In addition, we derive the density as a function of the
$UV$-velocity cutoff. The density (corrected for losses due to higher
$UV$-velocity cuts) starts to flatten out at $150\,\kms$ ($0.4\times
10^{-4}\,\ppc$), and is minimized (thus minimizing a possible non-halo
contamination) at $190\,\kms$ ($0.3\times 10^{-4}\,\ppc$). These
densities are in a rough agreement with the estimates for the {\it
stellar halo} white dwarfs, corresponding to a factor of 1.9 and 1.4
higher values.

\end{abstract}


\section{Introduction}

Dynamical studies indicate that the large quantities of gravitating
matter exist in the haloes of galaxies, including our own. The amount
of this matter far surpasses the matter with detectable
electromagnetic radiation. Revealing the nature of this so called
``dark matter'' is one of the crucial goals in present-day astronomy
and cosmology.

One of the first proposed methods to indirectly detect dark matter was
to look for lensing of background stars by the dark objects in the
halo \citep{pacz}. This technique, microlensing, is sensitive to
macroscopic objects, known as MACHOs, ranging in mass from planetary
to stellar. At the time when the first microlensing experiments began,
there was still no consensus as to whether the dark
matter haloes should be baryonic or non-baryonic. With the exception
of primordial black holes, microlensing would exclusively detect the
baryonic matter. Now, after years of observing, microlensing has ruled
out a dark halo made entirely of MACHOs, but has nonetheless found
more lensing events than expected from the known stellar populations,
either in the Galactic disk and halo, or in the LMC (or SMC), where
the lensed stars reside. The latest estimate from the MACHO group
favors a halo in which dark objects comprise 20\%, each with a typical
mass of $0.6 M_{\odot}$ \citep{alcock}. Another microlensing
experiment, EROS, puts upper limits for compact halo objects with
stellar masses of 30\%, from monitoring the LMC \citep{lasserre}, and
25\% from SMC \citep{afonso}. Some researchers (e.g., \citealt{sahu})
believe that lenses responsible for these events reside in the LMC (or
SMC) itself, however these scenarios have their own problems.

Since the search for dark matter in the form of MACHOs began, our
paradigms about what the cosmological dark matter should be have
changed. With the discovery of cosmological acceleration due to ``dark
energy'' \citep{sn1,sn2}, the discrepancy between the critical mass
density required by the inflationary model, and the low observed
densities of gravitating matter was reconciled. Cosmological models
preferred the genuine dark matter to be in the form of {\it
non-baryonic} cold dark matter, concentrated in galactic haloes. On
the other hand, the large fraction, or all of the {\it baryonic}
matter that was previously unaccounted for, is now believed to reside
in the systems such as those responsible for the population of Ly
$\alpha$ absorbers.

This alone should be enough to shift the focus on the detection of
non-baryonic halo dark matter (in the form of particles), were it not for
the MACHO results suggesting unknown, dark objects.  For the problem
to be more severe, there are a number of theoretical and observational
arguments suggesting that the baryonic matter should at best
constitute only a negligible fraction of the dark halo, much lower
than required to explain the fraction of it believed to be in MACHOs
\citep{graff}. Of the various candidate counterparts to MACHOs,
ancient, cool white dwarfs are nevertheless the best ones, especially
since their expected masses are comparable. The main question then,
is, whether we can detect these halo white dwarfs, and if so, what
fraction of the dark halo mass do they constitute. Ideally, these
white dwarfs should account for {\it all} MACHOs; otherwise yet
another form of baryonic dark matter is required.

Surveying 10\% of the sky, and using a proper-motion selected sample
of objects with high reduced proper motions, \citet{ohdhs} (hereafter
OHDHS) have identified and spectroscopically confirmed 98 white
dwarfs, of which 38 have halo-like kinematics based on the derived
velocity components in the plane of the Galaxy ($UV$ plane). The derived mass
density of these cool white dwarfs indicated that they make up at
least $\sim 2\%$ of the local dark-matter halo density, an order of
magnitude higher than expected from the population of the stellar halo
(as opposed to dark halo) white dwarfs (WDs). Stellar halo WDs should
differ from the dark halo ones in their origin, since the dark halo
white dwarfs were presumably produced in an early burst of star
formation, and were cooling ever since. The OHDHS paper and its
results stirred the astronomical community, generating numerous
objections -- regarding the selection of objects, kinematical cuts,
contamination from non-halo populations, and so on. A thorough review
on the subject of cool white dwarfs, and of various interpretations of
OHDHS work is given by \citet{araa}.

In this paper, we re-analyze the OHDHS sample of cool white dwarfs
using newly acquired high resolution spectra and CCD imaging. The new
data allow measurement of radial velocities and determination of
photometric distances in a more direct and precise manner, thus
addressing some issues raised with respect to original OHDHS
data. (The OHDHS used classification-grade, low resolution spectra,
and photometry derived from photographic plates).

In \S\ref{sec:spec} we present spectroscopic observations, and the
derivation of radial velocities for white dwarfs exhibiting spectral
features. Photometry from the CCD imaging is presented in
\S\ref{sec:phot}, together with a calibration of photographic
magnitudes used by OHDHS. In \S\ref{sec:data} we assemble all data to
produce a new dataset required for kinematical analysis. Finally, in
\S\ref{sec:disc} we discuss the properties of the revised kinematical
dataset, and use it to address some of the issues raised against the
original OHDHS sample and its interpretation. In particular, we
recalculate the densities of WDs using the OHDHS velocity cut, but
also as a function of the cutoff velocity. In a forthcoming work the
new dataset will be analyzed with additional techniques, including the
kinematical modeling of stellar populations.

\section{Spectroscopy \label{sec:spec}}

\subsection{Observations and data reduction}

The spectra of OHDHS cool white dwarfs were taken with the LRIS
\citep{oke} spectrograph on Keck I, on three nights: 2002 Sept 13, 14
and 2002 Dec 4 (UT). The primary goal was to obtain high resolution
spectra in the region around the \ha line ($\lambda 6563$). Thus, the
spectroscopic setup in the red arm of LRIS consisted of 1200/7500
grating, giving dispersion of 0.63 \AA\, ${\rm pix}^{-1}$, and a
spectral coverage of 5850--7140 \AA. The only exceptions to this setup
were in the cases of WDs with peculiar spectra (that happen to show no
hydrogen lines): WD2356-209\footnote{We use the same white dwarf
designations as in OHDHS} and LHS 1402. The first was imaged in both
the hi-res and the low-res modes, while the second only in the
low-res: 300/5000 grating, 2.46 \AA\, ${\rm pix}^{-1}$ dispersion, and
5010--10030 \AA\, range. The slit width of $1''$ gave the effective
resolution in hi-res mode of 2.9 \AA. On 2002 Dec 4 one additional WD
was observed (WD2346-478), this time with the 831/8200 grating (0.92
\AA\, ${\rm pix}^{-1}$ dispersion, 5630--7500 \AA\, range).

The cumulative exposure times varied from 6 to 45 minutes. Calibration
lamp spectra were obtained at each pointing, and the internal-lamp
flat-field images were taken once a night. Standard extraction and
calibration IRAF tasks were employed to produce the final spectra.

\subsection{Measuring radial velocities}

Measuring the radial velocities was the main goal of the hi-res
spectroscopy observing program. Targets were selected by inspecting the
low-res spectra obtained by OHDHS. They found an \ha line in 14 out of
38 cool WDs (denoted with an asterisk in OHDHS Table 1). Of these 14,
all but three can be observed from Keck's latitude, and we have
obtained spectra of all 11. In addition, we took spectra of another 12
WDs, thought to be featureless. Among these we find an additional two
with an \ha line. Thus our radial velocity sample comprises of 13 cool
WDs.

The wavelength region covered by the hi-res spectra would allow
detection of lines other than that of hydrogen (such as He and C). 
However, careful inspection of the spectra (with S/N ranging from $\sim 30$ 
to $\sim 110$) does not reveal any such lines.

Each individual spectrum was wavelength calibrated against a lamp
spectrum. The typical RMS of the calibration was 0.02 \AA. The
wavelengths of lines were first measured on individual spectra (of a
same object) in order to evaluate the stability of the zero point of
wavelength calibration, and to establish the accuracy with which the
central wavelength of \ha could be determined. All measurements were
done using the ``splot'' routine in IRAF. White dwarf spectra were
normalized by the continuum, and a Lorentzian function was used to
fit the lines.

The stability of the zero point of the calibration was determined by
measuring a bright sky emission line of [OI] at 6300.304 \AA. The
measured mean wavelength was $6300.30\pm0.01$ \AA, indicating no
systematic shifts in the calibration, while the scatter around the
mean of 0.07 \AA\, (equivalent to $3\,\kms$) gives the level of radial
velocity error due to the wavelength calibration.

The spectra of white dwarfs have different levels of signal to noise ratio,
leading to variations in the quality of the \ha profile. In well
exposed spectra, the non-LTE core of \ha line will be well defined, so
measuring the central wavelength of the core would be superior to
fitting a profile to the entire line. However, in lower S/N spectra
the core may be degraded by noise. In order to test which method is
more appropriate for our sample, we compared line fitting to the {\it
entire} profile (6400--6800 \AA) (which was fitting the wing portions
of the profile well, but often failed to fit the core), to the fitting
of the non-LTE core alone (width $\sim 5$ \AA). For each method, we
first find the difference of the individual measurements with respect
to the mean value (for the given object), and then calculate the
overall scatter. For wide-profile fitting the scatter is 0.45 \AA,
while for core fitting it is 0.32 \AA. In other words, core fitting
seems to have better repeatability, and should thus be more
precise. We should also note that we do not observe cases of split
cores or of emission in the cores.

Since the scatter, i.e., the error of the individual measurements, is
much larger than the stability of the zero point of wavelength
calibration, we conclude that it is safe to {\it combine} the spectra
belonging to the same object (between three and six), thus obtaining a
higher signal and eliminating deviant points by performing sigma
clipping. Since the spectra of a given object were taken over a short
period of time, the heliocentric velocity correction need not be
applied at this stage. We then proceed by finding central wavelengths
of \ha lines in the combined spectra, again by fitting a Lorentzian to
the core. This gives our final measured wavelength. In order to
evaluate the measurement error, we evaluate, as a function of a
measured flux, the RMS scatter of the central wavelengths of
individual spectra belonging to a given object. Not surprisingly, the
scatter is larger for objects whose individual spectra had low
intensities. We find a linear relation between the logarithm of flux
and the RMS scatter of individual wavelength measurements. However, at
a certain flux level the RMS reaches a minimum value of 0.13 \AA,
despite the increase of signal. In the fluxes of the {\it combined}
spectra this plateau is actually reached for most spectra in our
sample. Thus for most spectra, the total error from noise and
wavelength calibration uncertainties is equivalent to $\approx
7\,\kms$, well within the limits acceptable for this study. Finally,
to get the measured radial velocity, we apply the heliocentric
correction. The observed radial velocities of 13 WDs and their errors are
listed in Table \ref{tab:rv}. Also listed are 10 WDs observed with LRIS, 
but lacking spectral features.

\section{Photometry \label{sec:phot}}

\subsection{Observations and data reduction}

Photometry was performed on CCD images taken at the 1 m Nickel
Telescope at the Lick Observatory, on 2002 Nov 27, Dec 3 and 4
(UT). ``Dewar\#2'' CCD with a high quantum efficiency extending to
blue wavelengths was used. The first two nights were fully photometric
and many standards over a large range of airmasses and colors were
observed. This allowed construction of photometric transformations
with linear and quadratic color-terms. The photometric accuracy from
calibration is 0.01--0.02 mag in all bands.

Of the 38 cool WDs, 19 reach an altitude high enough to be observed
with this telescope. Of them, 18 were imaged in $V$ and $I$ (Cousins)
bands. In addition, 9 of those were also observed in $B$ band, and 
further 3 in $R$ band. Of the 18 WDs with $VI$ photometry, 9 also
belong to the radial velocity sample.

Object photometry was performed with an aperture equal to 1 FWHM of
the PSF (typically $2\farcs 4$), and then aperture-corrected using a
bright isolated star in the field. All $I$-band images were corrected
for fringing. For each measurement a photon error from the object was
combined with a photon error of the aperture-correction star. The
individual measurements were transformed into standard magnitudes and
combined (weighted by photometric errors) into a single magnitude per
star per band. Median photometry errors of the WD sample are
$\langle\sigma_B \rangle=0.053$, $\langle\sigma_V \rangle=0.035$,
$\langle\sigma_R \rangle=0.030$, $\langle\sigma_I
\rangle=0.035$. The median error of $V-I$ color, which we will use to
deduce distances, is 0.052 mag. The photometry is summarized in Table
\ref{tab:phot}.

Since most of the cool WDs were not known before, it is not surprising that
the literature search for photometric data produced prior measurements
for only two WDs: LHS 542 and LHS 147. The comparison of their
photometry to ours is given in Table \ref{tab:lit}. It is in excellent
agreement. Photometric magnitudes in other bands do exist for several
WDs in SDSS DR1, and for nine WDs in the 2MASS All-Sky Point Source
Catalog. The 2MASS measurements will be discussed in
\S\ref{ssec:color}.

\subsection{Photometric calibration of OHDHS magnitudes \label{ssec:calib}}

Since we obtained CCD photometry for only one half of the OHDHS
sample, it would be useful to derive photometry of other objects in
standard bands. Thus, we would like to construct empirical
transformations between the photographic plate magnitudes used by
OHDHS: $B_{\rm J}$, $R_{\rm 59F}$, and $I_{\rm N}$, and the standard
photometric bands. Empirical transformations between photographic and
standard magnitudes do exist in the literature, while the synthetic
transformations can be constructed using the model spectra and the
transmission curves, yet the first method has not been specifically
applied to stars such as cool WDs, while the second suffers from often
ill-defined properties of the actual response of a given plate/filter
combination.

Here we derive relations between photographic and standard
magnitudes as measured by CCD photometry.

For $B_{\rm J}$:
\be
B_{\rm J} = B - 0.85\, (B-V) + 0.26.
\ee

This relation has $\sigma = 0.10$. Since we know that $\sigma_B=0.05$,
this indicates that $\sigma_{B_{\rm J}}=0.08$, which is quite remarkable for
the photographic photometry. Note that a high color-term indicates
that (at least for WDs), $B_{\rm J}$ is actually closer to standard $V$ than
to standard $B$.

For $R_{\rm 59F}$:
\be
R_{\rm 59F} = V - 0.66\, (V-I) + 0.13. \label{eqn:vr59f}
\ee

Since we have our $R$ for only 3 objects, we give this transformation
relative to $V$. Excluded from the fit is LHS 1402 -- a peculiar
WD. Derived accuracy of the relation is 0.10 mag, while
$\sigma_{R_{\rm 59F}}=0.09$.

For $I_{\rm N}$:
\be
I_{\rm N} = I -0.09\, (V-I) + 0.12.
\ee

This relation has $\sigma\approx\sigma_{I_{\rm N}}=0.16$. Excluding objects
for which OHDHS derive $I_{\rm N}$ spectrophotometrically does not change
the above relation.

Another source of photographic magnitudes is the recently completed
USNO-B catalog (a similar catalog, GSC-2.2, does not go deep enough in
most cases). We have matched all the objects to counterparts in USNO-B
catalog \citep{usnob}, and repeated the above analysis against $B_2$,
$R_2$ and $I_{\rm SERC}$ -- second generation sky survey magnitudes
from USNO-B. However, we find that USNO-B magnitudes are significantly
inferior to those used by OHDHS, despite the fact that they come from 
similar or same plate material. Namely, we find $\sigma_{B_2}=0.41$,
$\sigma_{R_2}=0.58$, and $\sigma_{I_{\rm SERC}}=0.26$.

Overall, we conclude that the OHDHS (that is, SuperCOSMOS Sky Survey
from which it is taken, \citealt{sss1,sss2}) photographic plate
photometry is of excellent quality, which lends credence to
transforming them into standard magnitudes in order to derive
photometric distances.

Since we will be obtaining distances from $V$ magnitude and $V-I$
color, we want to directly transform OHDHS magnitudes and colors to
these. We have seen that $B_{\rm J}$ is quite close to $V$, so we use that 
magnitude to obtain the transformations

\be \label{eqn:calib1}
V = B_{\rm J} - 0.23\, (B_{\rm J} - I_{\rm N}) - 0.17, \qquad \sigma=0.12,
\ee
\be \label{eqn:calib2}
V-I = 0.62\, (B_{\rm J} - I_{\rm N}) - 0.04, \qquad \sigma=0.08.
\ee

These calibrations were derived omitting both peculiar-spectra WDs (LHS
1402 and WD2356-209).

\section{The new dataset \label{sec:data}}

\subsection{Radial velocities}

\subsubsection{Gravitational redshifts}

The observed radial velocities (Table \ref{tab:rv}) were extracted as
explained in \S\ref{sec:spec}. However, they do not represent the true
radial velocities, since WDs exhibit substantial gravitational redshift. 
The exact redshift depends on the mass and the radius of a
WD, which we do not know for individual WDs in our sample. However, it
is known that the range of these values is relatively small, so for
our purposes it is sufficient to adopt a common value for the
redshift. From \citet{reid} we find that the field WDs have an
average redshift of $28.6\,\kms$, with a spread of $6.5\,\kms$. We
subtract this value from the observed radial velocities, and add
the scatter to the radial velocity measurement error. The final values
are listed in Table \ref{tab:data}.

\subsubsection{Common proper motion binaries}

In a case where a WD has a common proper motion companion that is a
main sequence star, one can obtain a measurement of a true radial
velocity of a white dwarf (whether it contains spectral lines or not)
by simply measuring the radial velocity of the main sequence
component. Such a measurement circumvents the gravitational redshift
correction. To this end, we have carried out a search for companions
in the USNO-B catalog, which lists proper motions based on multiple
plates. Within the $2'$ search radius we find no candidate companions
with proper motions compatible to those of the white dwarfs.

\subsubsection{Selection effects \label{ssec:sel}}

The original OHDHS selection of cool WDs was based on the $U$ and $V$
components of the velocity. In the absence of radial velocities, they
were calculated by assuming $W=0$. Since our goal is to characterize
this population by obtaining the third component of the velocity from
the radial velocities, we should try to evaluate whether the subsample
for which radial velocities are measured is representative of the
population as a whole. Here we will restrict ourselves to a question
of whether the subsample is representative {\it kinematically} -- in
terms of its $U$ and $V$ velocities, based on which the cool WD sample
was selected in the first place.

OHDHS selected their sample by requiring the WDs to have a velocity
above some threshold in the $UV$-plane. This threshold was chosen as
$2 \sigma$ $UV$ velocity of the thick disk population

\be 
UV \equiv \sqrt{U^2+(V+35\,\kms)^2} > 94\,\kms.
\ee

One way of characterizing if the subsample of 13 WDs with radial
velocities is representative, is to compare its average $UV$ velocity
with the typical $UV$ velocities of randomly selected subsamples of 13
WDs out of the total 38.

\begin{figure}
\plotone{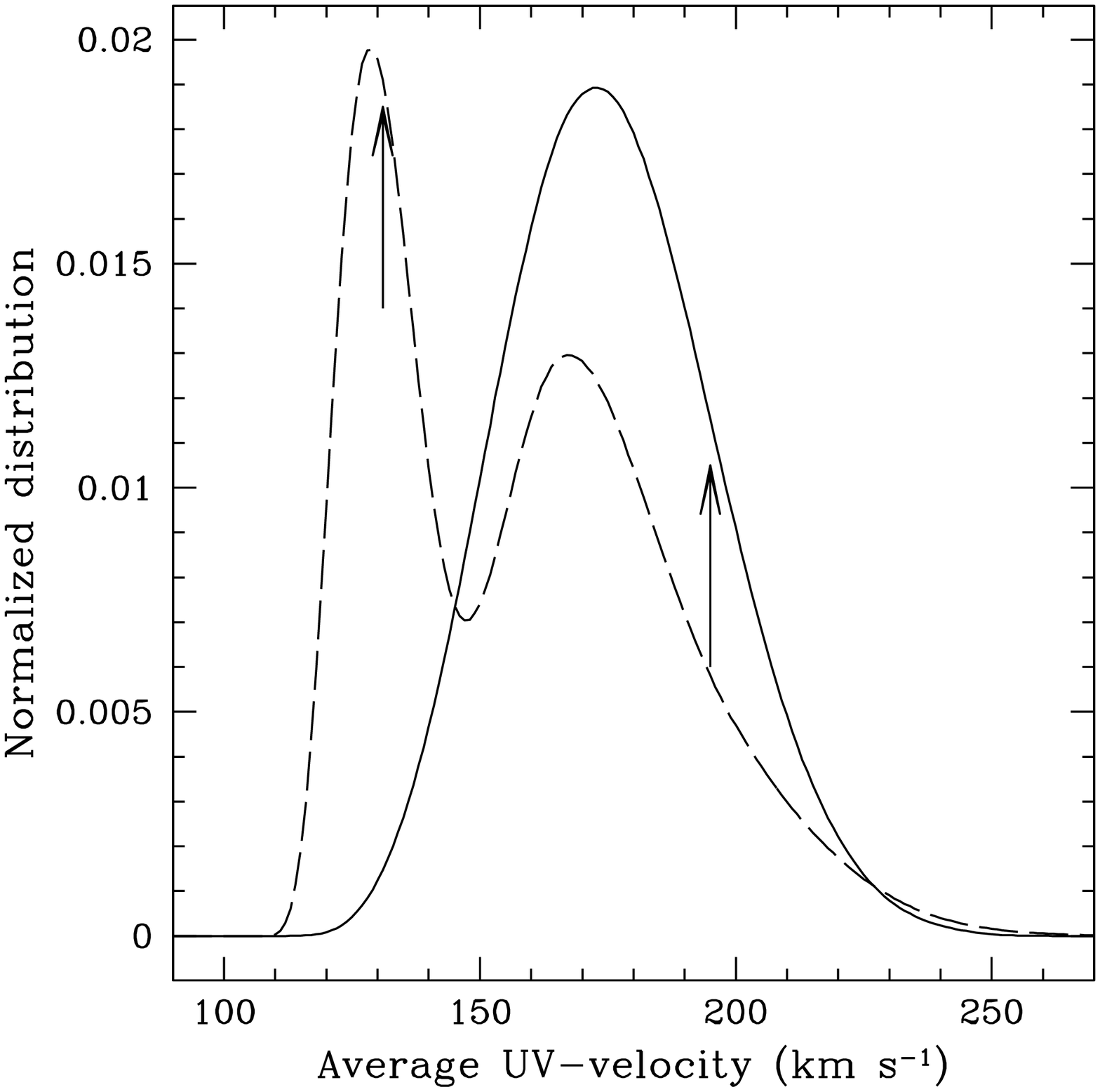}
\caption{Kinematical selection effects of the radial velocity
subsample. The two curves shows the straight (solid) and the weighted
(dashed) average $UV$-velocity ($UV\equiv \sqrt{U^2+(V+35\,\kms)^2}$)
of random subsamples of 13 white dwarfs (out of 38 OHDHS cool white
dwarfs). Compared to these distributions are the actual (straight and
weighted) $UV$-velocity averages of our radial velocity subsample,
shown with arrows.
\label{fig:uv_mc}}
\end{figure}

In order to obtain the distribution of $UV$ velocities of random
subsamples, we run a Monte Carlo simulation that draws 13 out of 38
WDs numerous times, and for each drawing calculates its average $UV$
velocity. The average is taken in two ways: as a straight average, and
as a weighted average, where weights are the corresponding maximal
volumes ($\langle UV \rangle = (\sum (UV/\vmax))/(\sum (1/\vmax))$) in
which a WD could have been detected in the OHDHS survey. As explained
in OHDHS, for each individual object, $\vmax$ is set either by the
magnitude limit of the survey, or by the lower proper motion cutoff,
whichever is smaller. In Figure \ref{fig:uv_mc}, the solid line
represents the distribution of unweighted averages. The unweighted
average of the radial velocity sample is $195\,\kms$, and is indicated
by the arrow. We see that it is on the high side of the distribution
(thus somewhat favoring fast objects), but well within the spread of
the distribution. The weighted $UV$ velocity distribution (dashed
line) has two peaks, two lower being dominated by proper-motion
limited objects, and the higher by the magnitude-limited ones. That we 
see two peaks might actually be indicative of the fact that
the OHDHS sample is composed of more than one population. We see that
the weighted average of our radial velocity sample lies right at the
proper-motion limited peak. If there really are two different
population, this might mean that our radial velocity subsample is
primarily representative of one of these populations -- the one with
intrinsically lower velocities. This will be further discussed in 
\S\ref{ssec:w}.

In any case, the radial velocity subsample does not seem to be extreme
with respect to the whole sample in terms of its $UV$-plane
kinematics.

\subsection{Proper motions}

Proper motions enter into the kinematical dataset since, together with
a distance, they determine two components of the physical
velocity. OHDHS proper motions come from SuperCOSMOS Sky Survey
\citep{sss3}. Since the OHDHS sample consists of relatively high
proper motion stars, the average fractional error is small (7\% from
listed values), and is thus not going to dominate in the velocity
error, especially since the distances were originally derived from
plate photometry. So, although not as important as other
recalibrations, we nevertheless carry out a comparison of SuperCOSMOS
proper motions of OHDHS WDs with those from the USNO-B catalog. USNO-B
combines a large number of plates to arrive at a proper motion
solution, the errors of which are found to be reliable
\citep{gould03}. In this comparison we use data from B.\ Oppenheimer's
online table\footnote{Available from
\url{http://research.amnh.org/users/bro}}, since unlike the published
version, it contains the individual components of the proper motion
error, just like the USNO-B catalog.

USNO-B contains proper motions for 36 OHDHS WDs. The median errors of
SuperCOSMOS proper motions are 3.5 times larger than those of
USNO-B. We find no systematic differences in two proper motion
datasets. The reduced $\chi^2$ between the two datasets is 0.8,
indicating good estimate of errors. (SuperCOSMOS proper motions were
recently also found by \citealt{digby} to agree with proper motions
derived by combining SuperCOSMOS and SDSS positions). There are four
cases in which either of the components is discrepant at $> 2 \sigma$
level. In all of these cases the listed error of USNO-B proper motions
is rather large, and also larger than the SuperCOSMOS listed
error. Visual inspection of DSS1 and DSS2 images confirms the
SuperCOSMOS value. Again, this is in line with \citet{gould03}, who
found that when USNO-B errors have large values, they are usually
underestimated. Thus, except in these four cases, and one other in
which USNO-B error is significantly larger than the SuperCOSMOS, in
Table \ref{tab:data} we mostly list USNO-B values, with a flag
indicating the source of proper motion.

\subsection{Distances}

\subsubsection{Colors and atmospheric composition \label{ssec:color}}

In principle, the multi-band photometry allows a determination of the
temperature of a white dwarf and its atmospheric composition. In
practice, we are often limited by the range of photometric
measurements and their precision. Nevertheless, construction of color-color
diagrams can be useful in some cases.

\begin{figure}
\plotone{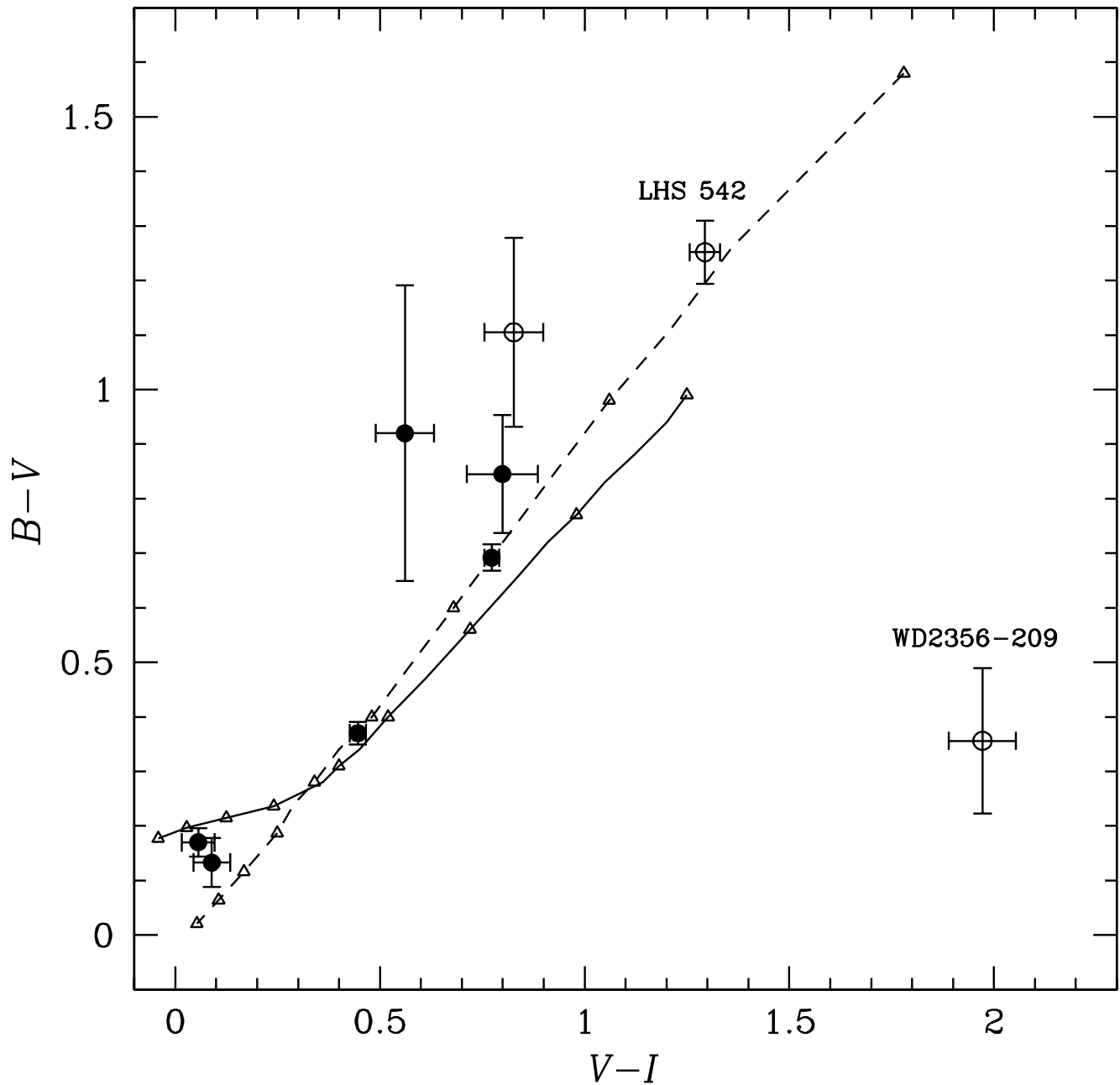}
\caption{$B-V$ vs. $V-I$ color-color diagram of OHDHS cool white dwarfs
with CCD photometry. Filled circles are white dwarfs exhibiting \ha
line (DA type).  The solid and the dashed tracks correspond to
theoretical colors for $g=8$ white dwarfs with pure hydrogen and pure
helium atmospheres, respectively. The models cover 4000--12,000 K
temperature range, with triangles marking every 1000 K. Labeled
objects are discussed in the text.
\label{fig:bvi}}
\end{figure}

From our CCD photometry we can place 9 OHDHS WDs onto a $BVI$ diagram
(Figure \ref{fig:bvi}). The solid and the dashed tracks correspond to
theoretical colors for $g=8$ white dwarfs with pure hydrogen and pure
helium atmospheres respectively, taken from \citet{berg95} and
\citet{berg95p}. Tracks go from 12,000 K on the blue end, to 4000
K. Filled symbols correspond to WDs showing \ha line (therefore of the
DA type). Judging from the position in the diagram, DA WDs are
consistent with the hydrogen track, as expected. Of the three non-DA
WDs, LHS 542 is clearly not consistent with a hydrogen atmosphere. As
shown by \citet{berg01}, a He atmosphere represents a good fit to LHS
542 optical and infrared photometry. More interesting is another
non-DA white dwarf, WD2356-209, an obvious outlier in the color-color
diagram. OHDHS have already shown its spectrum suggesting that it had
``no analogs''. Our LRIS spectra confirm this, and so does the
photometry -- we see excessively blue $B-V$ color for an extremely red
$V-I=1.97$.\footnote{Note that because $B_{\rm J}$ bandpass is
actually close to $V$, WD2356-209 did not stand out in the original
OHDHS color-color diagram (their Figure 4.)} It has been suggested
(I.\ N.\ Reid priv.\ comm.) that the heavy blanketing in the blue part
of the spectrum is due to an extremely broad Na I doublet (which would
make this WD a DZ type). Indeed, in our low-resolution spectra we see
a well defined dip around 5893 \AA. Thus the subdued flux in the Na I
region is consistent with $V-I$ being boosted, and $B-V$ becoming
blue. Recently a similar WD was found with an extremely wide Na I
absorption line -- SDSS J1330+6435 \citep{harris}. The S/N ratio of
SDSS J1330+6435 spectrum is too low to confirm the presence of other
lines characteristic of DZ WDs. Even in our 45 min low-res exposure we
cannot positively identify the Ca II triplet. It is possibly absent
because of a very low temperature.  Finally, the two hottest WDs in
our photometry sample are also in this diagram, and judging from their
position in it, they seem to have a temperature of $\sim 11,500$
K. Their $R$ magnitudes are also consistent with this temperature.

\begin{figure}
\plotone{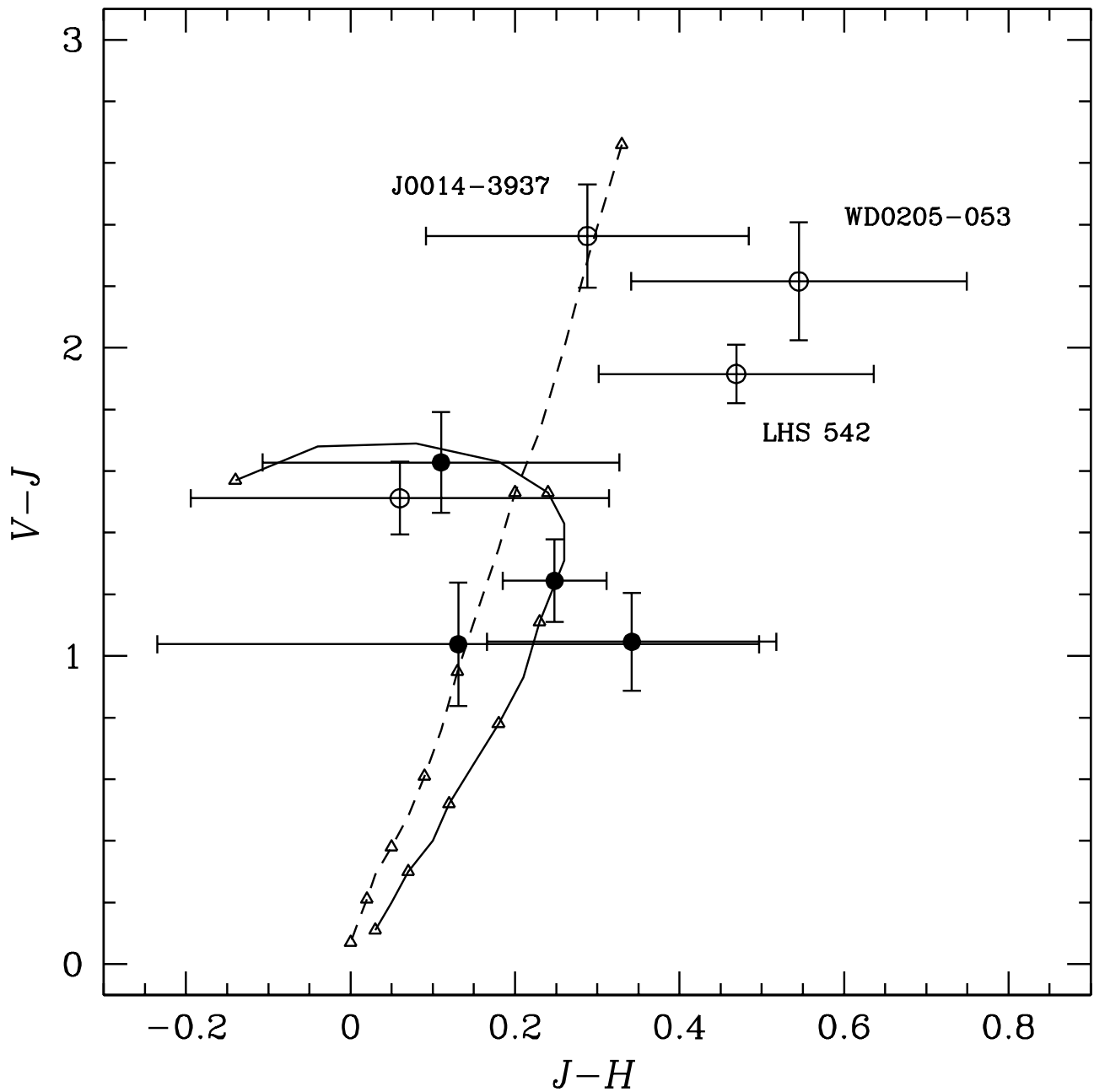}
\caption{$V-J$ vs. $J-H$ color-color diagram of OHDHS cool white dwarfs
present in 2MASS All-Sky Survey. See Figure \ref{fig:bvi} for
legend. Tracks terminate at 10,000 K.
\label{fig:vjh}}
\end{figure}

We also look for the OHDHS WDs in the 2MASS All-Sky Point Source Catalog. Nine
are catalogued. Since they are at the limits of 2MASS detection, the
infrared photometry is relatively crude. Actually, one is not
detected in $H$ band, and additional five lack $K_s$ magnitudes. Of
these, we have CCD photometry for only 3 (one of which is LHS 542,
discussed above). Therefore, for the remaining WDs we use a
calibration between $V$ and $B_{\rm J}$ (Eqn.\ \ref{eqn:calib1}),
whose accuracy is comparable to that of 2MASS magnitudes. We then plot
$V-J$ against $J-H$ in Figure \ref{fig:vjh}. Again, we can see that
all WDs with \ha are compatible with model pure hydrogen atmospheres
\citep{berg95}. Three do not seem to be consistent with H colors, of
which one is LHS 542. The other two (WD0205-053 and J0014-3937) are even 
cooler, with temperature $\sim 4200$ K.

Therefore, based on optical and infrared color information, we can
conclude that 4 WDs in the sample probably have He atmospheres: LHS 542,
WD2326-272, J0014-3937 and WD0205-053. In addition, \citet{berg03}, based on 
OHDHS photometry alone, finds that WD0125-043 and LHS 1447 are better fit 
with a He model.

Moreover, there are 3 additional WDs with $V-I>1.4$, a red color that
hydrogen WDs cannot attain. Finally, there are two WDs with peculiar
spectra (the previously mentioned WD2356-209 and LHS 1402), which will
be treated separately.

This leaves 11 cool WDs which, based on the above arguments, could
have either hydrogen or helium atmospheres. In order to use the
appropriate color-magnitude relation, we would like to classify these
remaining white dwarfs as well. \citet{berg97} have shown that all
hydrogen WDs with $T\gtrsim 5000$ K ($V-I \lesssim 1.0$) should
exhibit absorption lines. There are four such WDs without H lines, to
which we thus assign He atmospheres. We also see that of 6 WDs likely
to have He atmospheres based on color-color diagrams, all but LHS 1447
(He WD candidate based on \citealt{berg01} analysis of the original
OHDHS photometry) have $V-I \gtrsim 0.8$. Therefore, for the remaining
7 WDs (with $1.1<V-I<1.3$), we also adopt He composition, but will
allow for the possibility of misclassification.  Atmosphere
assignments are given in Table \ref{tab:data}.

\subsubsection{Color-magnitude relations \label{ssec:cmr}}

Currently, only LHS 542 has a good trigonometric parallax measurement,
with accuracy of 12\% (but see \ref{ssec:w}). Therefore, we have to
rely on photometric distances. It is to this end that we have acquired
precise $V-I$ CCD photometry. This was accomplished for half of the
sample. However, the CCD photometry also allowed the remaining WDs
with only photographic photometry to be transformed into standard
magnitudes (\S\ref{ssec:calib}). Note that the calibrations between
the plate and the standard magnitudes that appear in the literature
were not measured (or modeled) for white dwarfs (e.g,
\citealt{bessell,blair}), so their use might introduce systematic
offsets. On the other hand, our calibration is direct, so should be
free of {\it systematic} effects, and is accurate enough that it will
not dominate in the final distance error.

Next we need to decide what color-magnitude relation (CMR) to
use. Here we have a choice between using an empirically measured
relation (from WDs with trigonometric parallaxes), or model
relations.

The most widely used empirical CMR is the one based on \citet{berg01}
multi-band photometry of 152 WDs with measured parallaxes. The sample
is assembled from heterogeneous sources and is dominated by
disk WDs. A linear weighted fit to these WDs (of both DA and non-DA
type), produces a relation in $V-I$

\be \label{eqn:cmremp}
M_V = (2.72\pm0.09)\,(V-I) + (12.39\pm0.07)
\ee

This relation has a reduced $\chi^2=27$ indicating that we are
sampling a range of WD masses, or that some measurements are affected
by multiplicity.

OHDHS used \citet{berg01} empirical data to construct their CMR, and
then used model spectra with model bandpasses to convert standard
magnitudes into plate magnitudes.

In this paper, for deriving the distances, we will assume constant WD
mass of $0.6 M_{\odot}$ and thus use model cooling curves for hydrogen and
helium WDs. Surely, readers can use our photometry and a CMR of their
choice to arrive at different distance estimates.

For H atmospheres we use \citet{berg01} model CMR for WD
mass of $0.6 M_{\odot}$. Cooling tracks for other masses are
practically parallel, thus changing only the zero point of the
relation.

\be \label{eqn:cmrh} 
M_V = 3.42\,(V-I) + 11.7, \qquad (V-I<1.3) 
\ee

which is obviously steeper than the empirical relation given in Equation
\ref{eqn:cmremp}.

For helium atmospheres, again using the $0.6 M_{\odot}$ cooling curve of
\citet{berg01}, the following linear fit is appropriate for the color
range of interest

\be \label{eqn:cmrhe} 
M_V = 2.38\,(V-I) + 12.7, \qquad (V-I> 0.8).  
\ee

The helium relation, on the other hand, is somewhat less steep than
the empirical relation, at least for this red-color region. In our cool WD
sample there are two WDs with likely helium atmospheres that are bluer
than the above range. For them, we directly read values of $M_V$
from the He cooling curve.

As mentioned in the previous section, for 7 WDs we simply assume He
composition based on color.  If this assumption were not true, what
would be the error due to using a wrong CMR? For $V-I \lesssim
0.9$, H and He cooling tracks almost coincide, so there would be
almost no difference. For redder WDs, the error due to possible
misclassification is 

\be \label{eqn:sigmvcl}
\sigma_{M_V} ({\rm class}) = 1.10(V-I-1)+0.13 , \qquad (V-I > 0.9).
\ee

The CMRs given above were constructed for fixed masses of $0.6
M_{\odot}$. Various studies agree that this is a typical WD
mass. However, the spread of masses seems to be less well known, and
differs by as much as a factor of four (for a review see
\citealt{silv1}). To be conservative, we will assume a value close to
the higher estimates, $\sigma_M = 0.2 M_{\odot}$. This mass range
translates into absolute magnitude uncertainty of

\be \label{eqn:sigmv}
\sigma_{M_V} ({\rm H}) = 0.44; \qquad
\sigma_{M_V} ({\rm He}) = 0.50,
\ee

which we will use with the above color-magnitude relations.

\subsubsection{Peculiar white dwarfs}

As noted previously, OHDHS cool WD sample contains two objects with
peculiar properties. It would therefore be inappropriate to assign
$M_V$ to these objects based on $V-I$ color alone.

LHS 1402 has an extremely blue color of $V-I = -0.372$, (CCD
photometry).  Taken at face value, this would indicate an extremely
hot WD, with a temperature in excess of 100,000 K. Alternatively, this
could be a very cool hydrogen WD where the blue color is the result of
collision induced absorption by the ${\rm H_2}$ molecule
\citep{saumon1,hansen}. In a pure hydrogen model this would indicate a
temperature of only $\sim 2000$ K and $M_V$ of 18--19
\citep{saumon2}. However, \citet{berg02} have recently analyzed two
somewhat redder WDs with similar spectra, and concluded that strong
infrared suppression is better explained using a {\it mixed} H/He
model, where He dominates. Thus, using a mixed model with $N({\rm
H})/N({\rm He})=10^{-5}$, and $g=8$, for LHS 1402 we obtain $T=3000$ K
and $M_V=16.4$.  Note that LHS 1402 is discussed in \citet{berg03} as
possibly having a pure hydrogen atmosphere, and thus being extremely
faint and close. However, our LRIS spectra do not show a dip at $0.8
\micron$, suggesting that the mixed model is a better explanation. In
either case, this could well be the coolest white dwarf known.

Another WD with a peculiar spectral energy distribution, WD2356-209, has
already been discussed in terms of its photometry. One cannot use its
very red $V-I = 1.972$ to derive $M_V$ from CMR. It seems likely that its
temperature is in the 3500--4500 K range \citep{berg03}, and thus we
conservatively assign $M_V=16.5\pm 1.0$ to this object.

\subsubsection{Derived distances \label{ssec:dist}}

Finally, we use the appropriate CMRs to get absolute magnitudes and
thus the distance estimates from $V-I$ color -- both for the CCD
photometry sample, and for the sample with photometry calibrated from
plates, using Equations \ref{eqn:calib1} and \ref{eqn:calib2}. To get
a total error in absolute magnitude, we add in quadrature the error in
$M_V$ due to the uncertainty in $V-I$ color, the uncertainty due to a
possible range of WD masses (Eqn.\ \ref{eqn:sigmv}), and the
uncertainty due to possible misclassification, where appropriate
(Eqn.\ \ref{eqn:sigmvcl}). The error in $V$ magnitude is mostly negligible
and is correlated with the $V-I$ error, so we ignore it in calculating
the distance error. The resulting distances and their errors are
listed in Table \ref{tab:data}.

For LHS 542 we thus obtain a distance of $31\pm 7$ pc, in agreement
with the trigonometric parallax distance of $31.1\pm 3.6$ pc. Mostly
because of allowing for a large scatter in WD masses, our estimate of
a typical distance error is 24\%. If the spread in masses is actually
two times smaller ($0.1 M_{\odot}$), the distance error would also be
approximately cut in half. Our main goal, however, was to eliminate
possible {\it systematic} errors that would affect the kinematics, and
thus the interpretation of results. In the future, the parallaxes
should provide a definitive check.

Next, we compare the distances obtained here with those
listed in OHDHS. Taken together (but omitting the two peculiar WDs),
our new distances are 16\% {\it larger} than OHDHS (13\% if only WDs with
CCD photometry are considered). This difference is not just an overall
offset. Namely, for small distances, the two distance estimates agree
well, but the difference increases farther out, reaching on average
0.55 mag in distance modulus for the farthest stars (i.e., new
estimates are 30\% larger). Plotting the difference against $V-I$
shows a very similar trend -- new distances are larger for the blue
(almost exclusively hydrogen) white dwarfs. This is suggestive of the
fact that the difference arises from our use of a model CMRs, which
for hydrogen WDs (Eqn. \ref{eqn:cmrh}) has a steeper
slope than the empirical CMR (Eqn.\ \ref{eqn:cmremp}), a type
used by OHDHS. To test the significance of this effect, we
recalculated all distances using the {\it empirical} CMR (Eqn.\
\ref{eqn:cmremp}). The trend still exists, but is three times
smaller. 

\subsection{New cool white dwarf candidates \label{ssec:data}}

The OHDHS cool white dwarf sample of 38 stars was selected based on
their high velocities in the $UV$ plane. Does our recalibration of
distances make some white dwarfs exceed the velocity cutoff that they
previously were not able to reach? To answer this, we look at 60 WDs
that were identified by OHDHS, but had $UV\leqslant94\, \kms$. This
list was not published in OHDHS paper but is available online, at the
previously mentioned URL. Of 60, for 39 WDs both $B_{\rm J}$ and
$I_{\rm N}$ magnitudes are available, so we use Equations\
\ref{eqn:calib1} and \ref{eqn:calib2} to get $V$ and $V-I$. For the
remaining 21 WDs we obtain $V$ and $V-I$ from $B_{\rm J}$ and $R_{\rm
59F}$, again calibrated using our CCD photometry.  We then find
provisional distances according to hydrogen CMR (Eqn.\ \ref{eqn:cmrh})
for all objects with $V-I<0.8$, and the helium CMR (Eqn.\
\ref{eqn:cmrhe}) for red objects. Such assignment is conservative, so
that even if an incorrect CMR is used, the WD will not be excluded.

In this sample we notice two objects with the predicted distances of
only 7 pc. We identify one as LHS 69, a known nearby non-DA WD
\citep{berg01}. Its trigonometric parallax gives a distance of $8.1\pm
0.3$ pc. It has a standard $V=15.71$ (we predict 15.69). The second is
also a known WD with a mixed atmospheric composition \citep{berg94},
LHS 1126, at a trigonometric distance of $9.9\pm 1.0$ pc, and having
$V=14.50$ (predicted 14.37). Both of these cases provide additional
support that our calibration of plate photometry and the method for
estimating distances are not biased.

In this sample with new distances we find an additional 13 WDs,
previously falling below the velocity cutoff, to have $UV>94\,
\kms$. Predicted $UV$ velocities of these new cool WD candidates reach
as high as $157\, \kms$ in the case of LHS 4041. Actually, this white
dwarf, and one other of the thirteen (WD 0252-350) have actually
already been proposed, based on all three components of the velocity,
as candidate halo members \citep{spy}. Another six objects are also
listed in the literature -- of which five are spectroscopically
confirmed white dwarfs. The full data for these new candidate halo WDs
are given below the dividing line in Table
\ref{tab:data}.\footnote{Here we note that H.\ Harris (2003, priv.\ 
comm.) pointed out that one of the low velocity objects, WD0117-044,
is actually an almost equal WD binary, separated by $3''$, and has
measured for component A: $V=18.14\pm0.03$, $B-V=0.90\pm0.05$,
$V-I=1.00\pm0.03$, and for B: $V=18.17\pm0.03$, $B-V=0.95\pm0.05$,
$V-I=0.96\pm0.04$. This would make the distance estimate, and thus the
velocity, larger. However, using the USNO-B proper motion it still
falls somewhat short of the $94\, \kms$ cut.}

The effect goes in the opposite direction as well -- some of the
original OHDHS cool WDs no longer have estimated $UV$ velocities in
excess of $94\, \kms$. These will be discussed in the
\S\ref{ssec:dist_uv}.

Table \ref{tab:data} provides all data or measurements (with errors
when applicable) that are required for the kinematical
analysis. Besides the original 38 OHDHS cool WDs, we list 13 WDs that
qualify after the recalibration. For consistency with the original
paper, the names of previously unnamed WDs are constructed from OHDHS
J2000 coordinates.


\section{Discussion \label{sec:disc}}

\subsection{Radial velocities and the $UV$-plane velocities}

OHDHS selected their sample based on the two components of velocity
projected onto the sky. The third component, radial velocity, was not
available. In order to obtain components of motion in the Galactic
coordinate system ($UVW$), they assumed $W=0$, which produces some
arbitrary radial velocity that was then used to calculate $U$ and
$V$. Based on these $U$ and $V$ velocities they selected their cool WD
sample. Using the assumption of $W=0$, rather than $\vrad=0$, was seen
as a potential source of bias \citep{reid01,silv2}. The justification
given by OHDHS is based on the fact that their sample is mostly in the
direction of a Galactic pole, so $U$ and $V$ should not be much
affected by the $W$ component.

\begin{figure}
\plotone{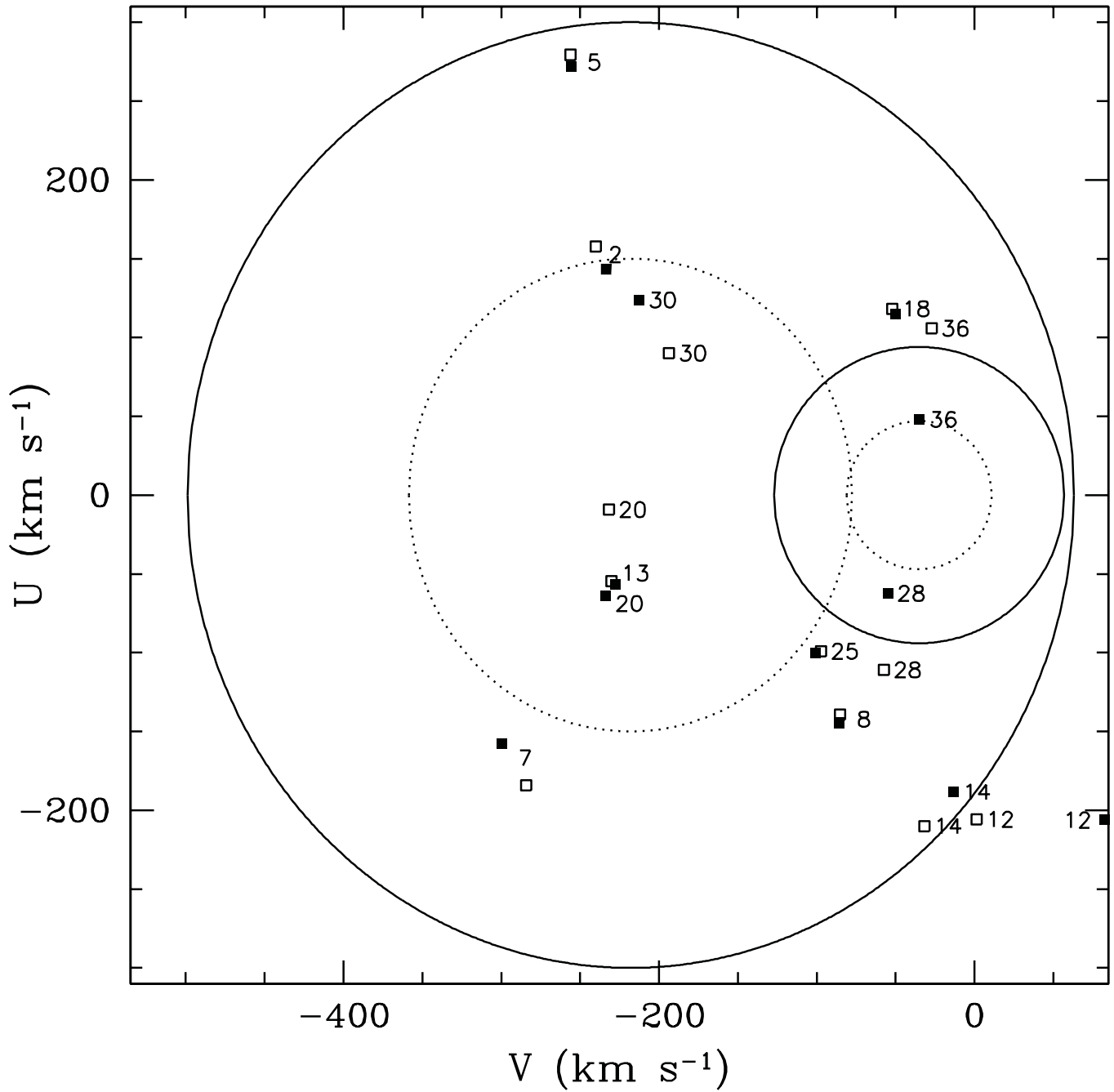}
\caption{The effect of radial velocities on the $UV$-plane
kinematics. We plot the original OHDHS positions of our radial
velocity sample (open squares), and their positions in the $UV$ plane
after the radial velocities have been take into account (filled
squares). Numbers should help match the corresponding points.
\label{fig:uv_rv}}
\end{figure}

For our radial velocity sample of OHDHS WDs we can calculate actual velocity components $U$, $V$ and $W$, and thus directly test and quantify the
validity of $W=0$ assumption. In Figure \ref{fig:uv_rv} we plot as
open squares the original\footnote{Note that here, and in the
entire paper, for LP 651-74 (line 36 in the OHDHS Table 1) we use
$B_{\rm J}-R_{\rm 59F}=0.72$, and thus the OHDHS distance of 39 pc, in
accordance with the online list.} OHDHS positions of 13 WDs with radial
velocities, and as filled squares their positions when the
radial velocities are take into account. To help match the
corresponding points we put sequential numbers next to them. We see
that the changes range from negligible to moderately high (point 12 =
LHS 4033 moved by $81\,\kms$). On average, the points have moved by
$29\,\kms$. What about the change in $UV$ velocity? First we see that
as a result of radial velocity knowledge, two objects that were just
outside of the $94 \,\kms$ cut have moved inward. These two cases
actually represent the largest changes ($-48$ and $-58\,\kms$), while
the average change is just $-5\,\kms$. On average, each individual
$UV$ velocity is smaller by 6\% (with a scatter of 21\%). Thus it
seems that this is a modest effect, and that the use of $W=0$
assumption in the absence of radial velocities is appropriate for this
sample. Note that in this comparison we kept the sky-projected
(tangential) velocities the same, i.e., we used the original OHDHS
values.

\subsection{New distances and the $UV$-plane velocities \label{ssec:dist_uv}}

Newly determined distances will, through modified sky-projected
velocities, directly affect the derived values of $U$ and $V$ velocity
components. We already saw the result of this in \S\ref{ssec:data},
where the new distances produced 13 new cool WD candidates with
potential halo kinematics. Also, in \S\ref{ssec:dist}, we saw that
there is a systematic trend affecting large distances more than the
nearby ones. Since on average we expect high $UV$ values to belong to
farther objects, we would expect that this trend will be reflected in
the $UV$ plane. The revised proper motions will also be responsible
for some change, albeit a very slight one.

\begin{figure}
\plotone{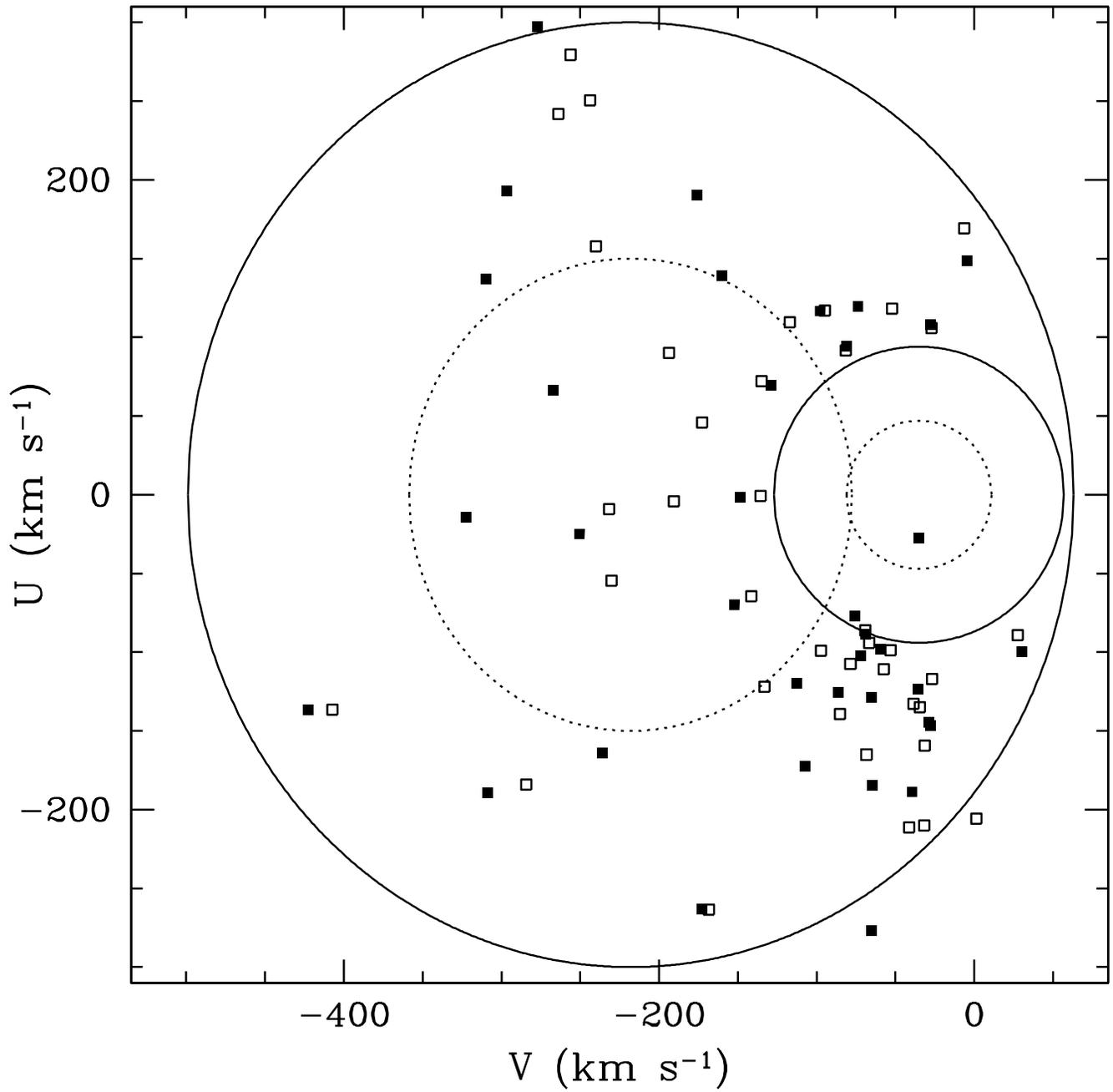}
\caption{The effect of recalibrated distances on the $UV$
velocities. Original velocities of 38 OHDHS cool white dwarfs are
shown as open squares (corresponding to their Figure 3), while values
obtained with new distances (and proper motions) are shown as filled
squares.
\label{fig:uv_dist}}
\end{figure}

In Figure \ref{fig:uv_dist} we show the new (filled squares), and the
original (open squares; equivalent to OHDHS Figure 3) $UV$-plane
positions of the 38 OHDHS WDs. In both cases the radial velocities are
neglected, i.e., $W=0$. To avoid clutter, individual points are not
labeled, yet we notice that the new $UV$ velocities tend to be higher,
especially for already high values. The analysis shows that on average
points have moved by $46\,\kms$, while the $UV$ velocity on average
increased by $23\,\kms$ (with maximum change being $+171\,\kms$). Each
individual $UV$ velocity is on average larger by 10\% (with a scatter
of 30\%).

Thus, the net effect of radial velocities and new distance
determinations is that the average $UV$ velocities are somewhat higher
than the original ones.

\subsection{Velocity component perpendicular to the Galactic plane 
\label{ssec:w}}

\begin{figure}
\plotone{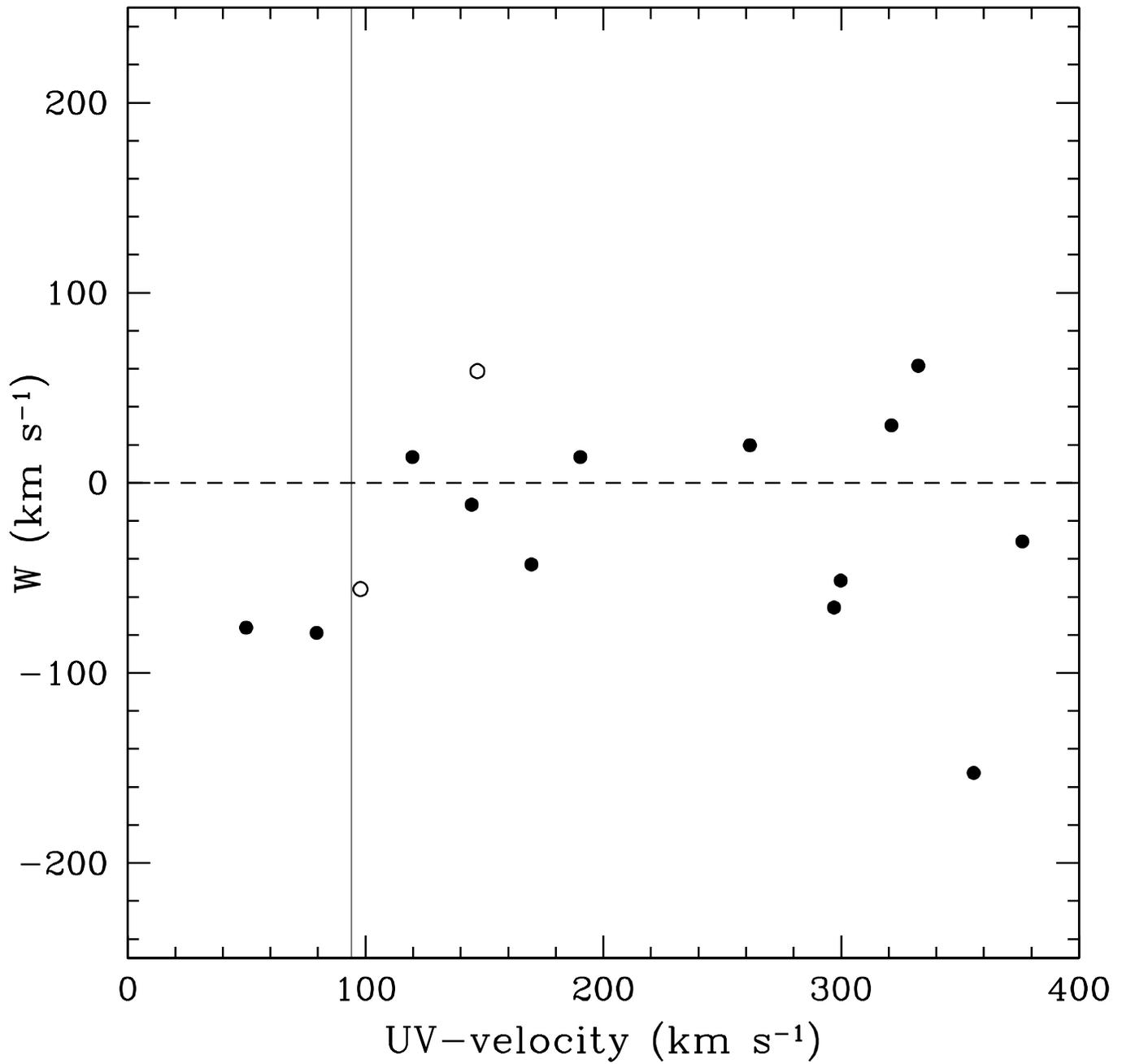}
\caption{Component of motion perpendicular to the Galactic plane. $W$
velocities of 13 OHDHS white dwarfs with radial velocities measured by
us are shown as filled circles. Two open circles come from the
``additional'' cool white dwarf candidates, and were measured by
\citet{spy}. Vertical line represents the $UV=94\,\kms$ cut.
\label{fig:w_uv}}
\end{figure}

For our radial velocity subsample of 13 WDs we can determine true
velocities in the direction perpendicular to the Galactic plane -- the
$W$ component. We plot the $W$ values in Figure \ref{fig:w_uv} as a
function of the $UV$ velocity. We have also added two WDs from the
newly qualified cool WDs that have radial velocities measured by
\citep{spy} (open circles). Omitting the two WDs with
$UV\leqslant94\,\kms$, we find the $W$ dispersion of the LRIS sample
to be $\sigma_W = 59\,\kms$. Only one WD (LHS 4033, $V-I=0.06$)
exceeds $100\,\kms$ -- reaching $W = -153\,\kms$.  Without it, we
would have $\sigma_W = 44\,\kms$. Actually, \citet{lhs4033} are about
to publish trigonometric parallax for this white dwarf. It turns out that
LHS 4033 is very massive ($M=1.25 M_{\odot}$), and thus some three
times closer than estimated based on $0.6M_{\odot}$ CMR. This measurement also
affects its surface gravity, producing much larger gravitational
redshift than we assume for our WDs. Altogether, if one was to use
this information a posteriori (which can be strongly argued against),
we would get $\sigma_W = 45\,\kms$. Anyhow, these values are in
between the values usually derived for thick disk and spheroid (halo)
populations, $35\,\kms$ and $94\,\kms$, respectively \citep{cb}. That
the radial velocity sample probes mostly what appears to be a lower
velocity (and younger?)  population was already indicated in
\S\ref{ssec:sel}. Indeed, all but three of the radial velocity WDs are
proper-motion limited. While one can formally calculate $\sigma_U$ and
$\sigma_V$, they are meaningless because of the $UV$ selection that
was applied to derive the sample in the first place. We defer a more
thorough analysis of the kinematics and population analysis to a
forthcoming paper.

\subsection{Space density of white dwarfs}

The derived space density, and thus the mass density, of OHDHS cool
white dwarfs was their key result, and the one that stirred most
controversy, since it is considerably higher than the expected stellar (as
opposed to dark matter) halo WD density. Using OHDHS original data for 38
cool WDs, and applying the same $1/\vmax$ technique, with a limiting
magnitude of $R_{\rm 59F}^{\rm lim} = 19.8$, one derives $n = 1.4
\times 10^{-4}\,\ppc$ (close to the value given by OHDHS of $1.8
\times 10^{-4}\,\ppc$). Using a typical WD mass of $0.6 M_{\odot}$,
this corresponds to $\rho = 8.3 \times 10^{-5}\,\mpc$, or some six
times higher than the canonical value of stellar halo WD mass density
of $\rho_c = 1.3 \times 10^{-5}\,\mpc$ \citep{gfb}. (Note that this
often quoted {\it canonical} value is somewhat of an educated guess,
and not a real measurement (A.\ Gould, private communication 2003)).

What estimate of density would we get with our updated data? In our
analysis we include all cool WD candidates from Table \ref{tab:data},
using the radial velocity data where available. For completeness, we
append this list with the remaining low-velocity WDs identified by
OHDHS ($UV\leqslant94\,\kms$). For them, we use listed SuperCOSMOS
proper motions and do not check for the availability of radial
velocities in the literature. Looking at the entire WD sample of 98
objects will allow us to characterize the density not just for a
sample with a fixed $UV$ cut, but rather as a function of the
$UV$-cutoff velocity.

Of 47 WDs with $UV>94\,\kms$ only 12 are magnitude-limited. All
others, including all 51 WDs with $UV\leqslant94\,\kms$ are proper-motion
limited, that is, the maximum distance at which they could be detected
is determined by the proper-motion lower limit of the survey ($330\,
\masyr$). For the magnitude limit we use $R_{\rm 59F}^{\rm lim}= 19.8$
transformed for each object into a corresponding $V^{\rm lim}$ using 
the Equation \ref{eqn:vr59f}.

If we restrict ourselves to the original $UV$-cut of $94\,\kms$, the 47
WDs that make this cut yield
\be
n_{94} = 1.72\times 10^{-4}\,\ppc
\ee

which is similar or slightly higher than the original estimate based
on 38 WDs and $R_{\rm 59F}^{\rm lim} = 19.8$. For this sample we have
$\langle {\mathbf V}/\vmax \rangle = 0.51$, suggesting that
$R_{\rm 59F}^{\rm lim}$ has an appropriate value.

\begin{figure}

\plotone{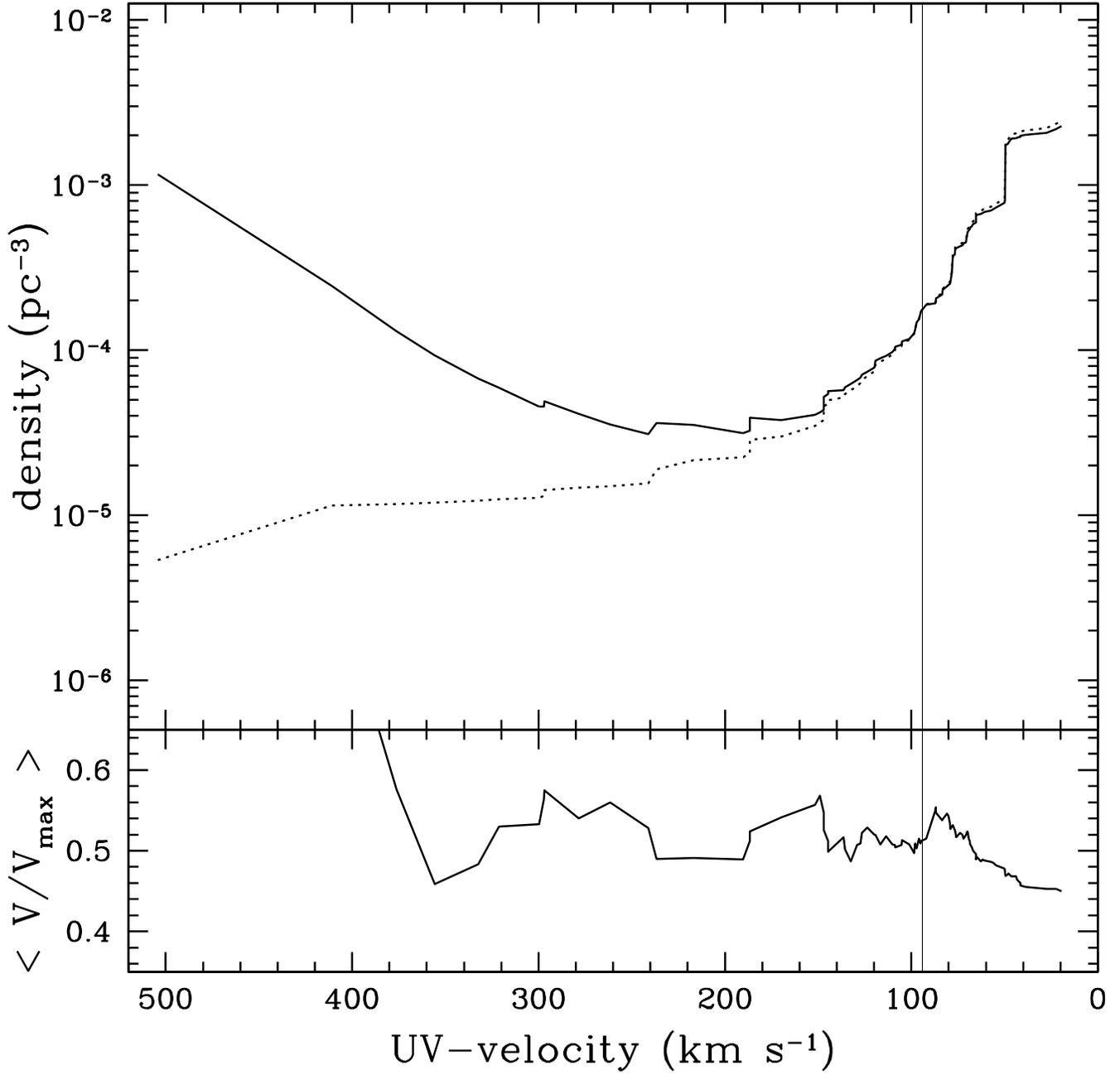}
\caption{Number density of OHDHS white dwarfs as a function of $UV$-cut
velocity. In the upper panel we show the corrected (solid line) and
the uncorrected (dotted line) cumulative density of all 98 white
dwarfs observed by OHDHS (summing from the high-velocity end). The
corrected line takes into account that at high $UV$ values we sample a
smaller part of the $UV$ plane. Vertical line represents the
$UV=94\,\kms$ cut. In the lower panel we monitor average $\langle
{\mathbf V}/\vmax \rangle$ as we move towards the lower $UV$
velocities. Note that beyond $\sim 300\,\kms$ we only have a couple of
objects.
\label{fig:dens}}
\end{figure}

How sensitive is this estimate on the choice of $UV$-cutoff velocity?
In Figure \ref{fig:dens} we plot cumulative number density starting
from the {\it highest} $UV$ values (dotted line). However, since at
higher $UV$ values we limit ourselves to yet smaller portion of the
$UV$-plane, we need to correct it. We do it by finding (at each $UV$)
a fraction of objects with \citet{cb} halo kinematics that get
excluded by different $UV$ cuts. (The correction factor is normalized
to 1 for $UV=94\,\kms$, in order to make the results directly
comparable to OHDHS who did not perform this correction. In any case,
the actual correction at this velocity is very small).

The corrected densities are shown with a solid line. The lower panel
monitors $\langle {\mathbf V}/\vmax \rangle$ at each point, and
appears consistent with 0.5 for the entire range of interest. The
vertical thin line denotes the $94\,\kms$ limit. Inward of this limit
we have a rise of density due to non-halo populations. Actually, we see
that this rise begins inward of $UV\sim 150\,\kms$. Note that in the
{\it corrected} plot, the density of objects with halo kinematics
should be independent of the $UV$ cut. The corrected density at
$UV=150\,\kms$ is
\be
n_{150} = 0.42\times 10^{-4}\,\ppc,
\ee
corresponding to $1.9\,\rho_c$. The minimum value of the attained density is
\be
n_{\rm min} = 0.31\times 10^{-4}\,\ppc,
\ee
at $UV\sim 190\,\kms$. This minimum value corresponds to $1.4\,\rho_c$
-- within the uncertainty of the stellar halo WD density.\footnote{Note that 
\citet{lhs4033} result on LHS 4033 would only slightly affect $n_{94}$ (increasing it by 5\%), and would leave $n_{150}$ and $n_{\rm min}$ unchanged.}
Beyond $UV\sim 250\,\kms$, the density estimate rises again. It exceeds
$n_{94}$ when $UV>400\,\kms$. However, in this range, the density is based 
on just 2--3 highest velocity objects with huge correction factors.

\section{Conclusions}

We obtain precise radial velocities for the majority of OHDHS WDs with
\ha lines. This makes it possible to measure more precisely the $U$ and $V$
components of the velocity, and also allows the $W$ component to be
derived. We show that the radial velocities do not affect
significantly the way in which OHDHS selected their cool WD
candidates. Our $W$-velocity dispersion lies between the typical thick
disk and halo values, an indication of a mixed sample. 

Also, our new CCD photometry, and the recalibration of OHDHS SuperCOSMOS
plate photometry, allow for more robust distance estimates.

Finally, with the new dataset, and applying the same methods of
analysis as in OHDHS, we confirm the densities of cool white dwarfs
that they derived. However, many times lower densities (consistent
with the stellar halo) are found if one adopts higher $UV$-cutoff 
velocities. This new set of data facilitates a more sophisticated
analysis, which we plan to present in a forthcoming paper.

\acknowledgments We thank Didier Saumon for valuable discussions, and
the referee Hugh Harris for sharing some unpublished measurements. We
are indebted to Nigel Hambly and Andrew Digby for their work in
constructing the SuperCOSMOS sample.
This publication makes use of VizieR and SIMBAD Catalogue Services of CDS
in Strasbourg, France, and data products from the Two Micron All Sky
Survey, which is a joint project of the University of Massachusetts
and the IPAC/Caltech, funded by the NASA and the NSF.

\clearpage


\begin{deluxetable}{llrr}
\tabletypesize{\footnotesize}
\tablecolumns{2} 
\tablewidth{0pt} 
\tablecaption{Observed WDs and their radial velocities.\tablenotemark{a} 
\label{tab:rv}} 
\tablehead{ No.\tablenotemark{b} & Name & $v^{\rm obs}_{\rm rad}$ ($\kms$)}
\startdata 
2       &       WD0153-014     &      $-50 \pm     7$ \\
5       &       LHS 147        &      $ -15   \pm   7$ \\
7       &       WD0135-039     &      $ -28   \pm   7$ \\
8       &       LHS 4042       &      $ -24   \pm   7$ \\
9       &       WD2356-209     &       \nodata \\
10      &       WD0227-444     &       \nodata \\
12      &       LHS 4033       &      $ 206   \pm   7$ \\
13      &       LP 586-51      &      $ -22   \pm   7$ \\
14      &       WD2242-197     &      $ -8   \pm    7$ \\
15      &       WD0205-053     &       \nodata \\
17      &       WD0125-043     &       \nodata \\
18      &       WD2346-478     &      $ 75   \pm    7$ \\
20      &       WD0300-044     &      $ 141  \pm    13$ \\
21      &       WD0123-278     &       \nodata \\
24      &       LHS 1402       &       \nodata \\
25      &       LHS 1274       &      $ 81   \pm    7$ \\
27      &       WD0044-284     &       \nodata \\
28      &       WD2214-390     &      $ 50   \pm    7$ \\
30      &       LP 588-37      &      $ -127  \pm   7$ \\
32      &       WD0045-061     &       \nodata \\
33      &       WD0225-326     &       \nodata \\
35      &       WD0117-268     &       \nodata \\
36      &       LP 651-74      &      $ 74   \pm    7$ \\
\enddata
\tablenotetext{a}{In cases where no features were detected, radial 
velocity is omitted}
\tablenotetext{b}{Numbering follows Table 1 of OHDHS}
\end{deluxetable}

\begin{deluxetable}{llrrrrr}
\tabletypesize{\footnotesize}
\tablecolumns{6}
\tablewidth{0pt} 
\tablecaption{Johnson-Cousins CCD photometry \label{tab:phot}} 
\tablehead{ No.\tablenotemark{a} & Name & $B$ & $V$ & $R$ & $I$ & 
$n_{\rm obs}$\tablenotemark{b}}
\startdata 
2 & WD0153-014  &  \nodata          &  $18.646\pm0.020$ &   \nodata         &  $18.415\pm0.034$ &  0202  \\
3 & LHS 542      &  $19.473\pm0.053$ &  $18.221\pm0.023$ &  $17.545\pm0.025$ &  $16.926\pm0.028$ &  2321  \\
5 & LHS 147     &  $17.985\pm0.018$ &  $17.615\pm0.011$ &   \nodata         &  $17.169\pm0.016$ &  3202  \\
6 & WD2326-272   &  $21.027\pm0.165$ &  $19.922\pm0.051$ &   \nodata         &  $19.095\pm0.051$ &  2202  \\
7 & WD0135-039  &   \nodata         &  $19.644\pm0.048$ &  \nodata          &  $19.083\pm0.065$ &  0402  \\
9 & WD2356-209   &  $21.206\pm0.109$ &  $20.850\pm0.075$ &   \nodata         &  $18.878\pm0.033$ &  2404  \\
12 & LHS 4033    &  $17.162\pm0.020$ &  $16.992\pm0.017$ &  $16.987\pm0.030$ &  $16.936\pm0.036$ &  3222  \\
13 & LP 586-51  &  $18.318\pm0.039$ &  $18.185\pm0.022$ &  $18.141\pm0.044$ &  $18.096\pm0.039$ &  1212  \\
14 & WD2242-197  &  $20.504\pm0.098$ &  $19.659\pm0.045$ &   \nodata         &  $18.861\pm0.074$ &  1201  \\
15 & WD0205-053  &   \nodata         &  $18.898\pm0.145$ &  \nodata          &  $17.257\pm0.024$ &  0101  \\
17 & WD0125-043  &   \nodata         &  $19.820\pm0.076$ & \nodata           &  $18.911\pm0.054$ &  0202  \\
20 & WD0300-044  &  $20.782\pm0.266$ &  $19.862\pm0.049$ &   \nodata         &  $19.301\pm0.051$ &  1404  \\
24 & LHS 1402    &   \nodata         &  $18.050\pm0.014$ &  \nodata          &  $18.422\pm0.027$ &  0205  \\
27 & WD0044-284  &   \nodata         &  $20.022\pm0.060$ &  \nodata          &  $18.713\pm0.039$ &  0402  \\
30 & LP 588-37  &  \nodata          &  $18.496\pm0.024$ &  \nodata          &  $18.365\pm0.039$ &  0303  \\
32 & WD0045-061  &  \nodata          &  $18.203\pm0.015$ &  \nodata          &  $17.219\pm0.019$ &  0302  \\
35 & WD0117-268  &  \nodata          &  $19.057\pm0.049$ &  \nodata          &  $17.944\pm0.030$ &  0303  \\
36 & LP 651-74  &  $18.033\pm0.021$ &  $17.342\pm0.011$ &   \nodata         &  $16.568\pm0.015$ &  2202  \\
\enddata
\tablenotetext{a}{Numbering follows Table 1 of OHDHS}
\tablenotetext{b}{Number of observations for each band: $BVRI$}
\end{deluxetable} 

\begin{deluxetable}{lrrrrl}
\tabletypesize{\footnotesize}
\tablecolumns{6} 
\tablewidth{0pt} 
\tablecaption{Comparison with the published photometry \label{tab:lit}}
\tablehead{ Name & $B$ & $V$ & $R$ & $I$ & Reference }
\startdata
LHS 542  & 19.23 & 18.15 & 17.53   &  16.99  & \citet{berg01} \\
         & 19.49 & 18.25 & \nodata & \nodata & \citet{ldm} \\
         & 19.47 & 18.22 & 17.55   &  16.93  &  This work \\
& & & & & \\
LHS 147 & 17.97 & 17.62 & 17.38   &  17.16  & \citet{berg97} \\
         & 18.09 & 17.66 & \nodata & \nodata & \citet{ldm} \\
         & 17.94 & 17.57 & \nodata & \nodata & \citet{es} \\
         & 17.99 & 17.62 & \nodata &  17.17  & This work \\ 
\enddata
\end{deluxetable} 

\begin{deluxetable}{llrrrrrrlrrlrrlrr}
\tabletypesize{\scriptsize}
\tablecolumns{17} 
\tablewidth{0pt} 
\tablecaption{The new dataset -- relevant data on cool white dwarfs \label{tab:data}}
\tablehead{ No. & Name & R.A.   & Dec. & 
$\mu_{\alpha}$ & $\mu_{\delta}$ & $\sigma(\mu_{\alpha})$ & 
$\sigma(\mu_{\delta})$ & Flag A\tablenotemark{a} & $V$ & $V-I$ & 
Flag P\tablenotemark{b} & $d$  & $\sigma_d$ & Flag C\tablenotemark{c} &
$v_{\rm rad}$ & $\sigma(v_{\rm rad})$ \\
               &      & (deg)  & (deg) &
$\masyr$       & $\masyr$       & $\masyr$               &
  $\masyr$             &                         & mag & mag   &
                        & pc   &  pc        &                         &
$\kms$        & $\kms$ }
\startdata
1 & F351-50 & 11.33178 & $-$33.49130 & 1860 & $-$1486 & 51 & 10 & U & 19.37 & 1.54 & O & 40 & 10 & He & \nodata & \nodata \\
2 & WD0153-014 & 28.46448 & $-$1.39468 & 64 & $-$398 & 4 & 6 & U & 18.65 & 0.23 & S & 170 & 36 & H & $-$79 & 9 \\
3 & LHS 542 & 349.78956 & $-$6.21383 & $-$618 & $-$1584 & 1 & 5 & U & 18.22 & 1.29 & S & 31 & 7 & He & \nodata & \nodata \\
4 & WD0351-564 & 57.78907 & $-$56.45198 & 265 & $-$1052 & 20 & 19 & O & 20.96 & 1.49 & O & 88 & 22 & He & \nodata & \nodata \\
5 & LHS 147 & 27.03805 & $-$17.20401 & $-$120 & $-$1106 & 7 & 6 & U & 17.61 & 0.45 & S & 75 & 15 & H & $-$44 & 9 \\
6 & WD2326-272 & 351.54458 & $-$27.24632 & 576 & $-$104 & 4 & 9 & U & 19.92 & 0.83 & S & 112 & 27 & He & \nodata & \nodata \\
7 & WD0135-039 & 23.89029 & $-$3.95502 & 454 & $-$186 & 7 & 6 & U & 19.64 & 0.56 & S & 160 & 38 & H & $-$57 & 9 \\
8 & LHS 4042 & 358.64586 & $-$32.35540 & 422 & $-$46 & 1 & 10 & U & 17.41 & 0.18 & O & 105 & 25 & H & $-$52 & 9 \\
9 & WD2356-209 & 359.18788 & $-$20.91370 & $-$329 & $-$211 & 32 & 20 & O & 20.85 & 1.97 & S & 74 & 34 & sp & \nodata & \nodata \\
10 & WD0227-444 & 36.87318 & $-$44.38573 & 268 & $-$217 & 12 & 18 & O & 19.82 & 1.06 & O & 83 & 22 & He? & \nodata & \nodata \\
11 & J0014-3937 & 3.44777 & $-$39.62331 & $-$226 & $-$714 & 17 & 2 & U & 18.70 & 1.25 & O & 40 & 10 & He & \nodata & \nodata \\
12 & LHS 4033\tablenotemark{d} & 358.13289 & $-$2.88647 & 614 & 324 & 10 & 8 & U & 16.99 & 0.06 & S & 105 & 22 & H & 178 & 9 \\
13 & LP 586-51 & 15.53001 & $-$0.54986 & 342 & $-$122 & 3 & 3 & U & 18.19 & 0.09 & S & 172 & 37 & H & $-$51 & 9 \\
14 & WD2242-197 & 340.43428 & $-$19.67841 & 346 & 62 & 4 & 4 & U & 19.66 & 0.80 & S & 111 & 27 & H & $-$36 & 9 \\
15 & WD0205-053 & 31.29830 & $-$5.29836 & 956 & 400 & 3 & 6 & U & 18.90 & 1.64 & S & 29 & 8 & He & \nodata & \nodata \\
16 & WD0100-645 & 15.20987 & $-$64.48649 & 516 & 190 & 5 & 0 & U & 17.58 & 0.58 & O & 60 & 14 & H & \nodata & \nodata \\
17 & WD0125-043 & 21.27431 & $-$4.28424 & 250 & $-$44 & 6 & 4 & U & 19.82 & 0.91 & S & 98 & 25 & He & \nodata & \nodata \\
18 & WD2346-478 & 356.51213 & $-$47.85060 & $-$270 & $-$454 & 5 & 3 & U & 17.95 & 0.83 & O & 48 & 11 & H & 47 & 10 \\
19 & LHS 1447 & 42.05496 & $-$30.02575 & 436 & 322 & 0 & 6 & U & 18.43 & 0.44 & O & 106 & 25 & He & \nodata & \nodata \\
20 & WD0300-044 & 45.09852 & $-$4.42355 & 272 & $-$280 & 17 & 19 & O & 19.86 & 0.56 & S & 177 & 41 & H & 112 & 15 \\
21 & WD0123-278 & 20.76574 & $-$27.80398 & 342 & 124 & 5 & 16 & U & 20.29 & 1.29 & O & 80 & 26 & He? & \nodata & \nodata \\
22 & WD2259-465 & 344.77772 & $-$46.46632 & 404 & $-$158 & 4 & 7 & U & 19.71 & 1.26 & O & 64 & 20 & He? & \nodata & \nodata \\
23 & WD0340-330 & 55.03620 & $-$33.01671 & 494 & $-$330 & 3 & 5 & U & 19.94 & 1.17 & O & 77 & 22 & He? & \nodata & \nodata \\
24 & LHS 1402 & 36.13432 & $-$28.91646 & 492 & $-$30 & 3 & 2 & U & 18.05 & $-$0.37 & S & 21 & 5 & sp & \nodata & \nodata \\
25 & LHS 1274 & 24.80995 & $-$33.81756 & 580 & $-$24 & 3 & 11 & U & 17.34 & 0.49 & O & 62 & 15 & H & 52 & 9 \\
26 & WD0214-419 & 33.56203 & $-$41.85251 & 320 & $-$96 & 18 & 19 & O & 20.08 & 0.98 & O & 102 & 25 & He & \nodata & \nodata \\
27 & WD0044-284 & 11.00892 & $-$28.40313 & $-$78 & $-$360 & 13 & 3 & U & 20.02 & 1.31 & S & 69 & 23 & He? & \nodata & \nodata \\
28 & WD2214-390 & 333.64480 & $-$38.98522 & 1006 & $-$360 & 2 & 8 & U & 16.14 & 0.67 & O & 27 & 6 & H & 21 & 9 \\
29 & WD2324-595 & 351.04227 & $-$59.46895 & 124 & $-$576 & 5 & 7 & U & 16.90 & 0.14 & O & 88 & 21 & H & \nodata & \nodata \\
30 & LP 588-37 & 25.58649 & $-$1.39757 & 108 & $-$344 & 1 & 7 & U & 18.50 & 0.13 & S & 186 & 40 & H & $-$155 & 9 \\
31 & WD0345-362 & 56.38631 & $-$36.18446 & 142 & $-$588 & 18 & 67 & U & 20.40 & 1.45 & O & 71 & 17 & He & \nodata & \nodata \\
32 & WD0045-061 & 11.27623 & $-$6.13876 & 104 & $-$676 & 3 & 3 & U & 18.20 & 0.98 & S & 43 & 10 & He & \nodata & \nodata \\
33 & WD0225-326 & 36.36950 & $-$32.63163 & 310 & 160 & 36 & 4 & U & 18.61 & 0.40 & O & 118 & 28 & He & \nodata & \nodata \\
34 & WD2348-548 & 357.19527 & $-$54.76280 & 364 & $-$96 & 22 & 32 & U & 19.21 & 0.98 & O & 69 & 17 & He & \nodata & \nodata \\
35 & WD0117-268 & 19.46521 & $-$26.81428 & 476 & 42 & 3 & 3 & U & 19.06 & 1.11 & S & 55 & 15 & He? & \nodata & \nodata \\
36 & LP 651-74 & 46.80880 & $-$7.24976 & $-$193 & $-$436 & 11 & 10 & O & 17.34 & 0.77 & S & 40 & 8 & H & 45 & 9 \\
37 & WD0135-546 & 23.91108 & $-$54.59108 & 660 & 108 & 17 & 3 & U & 18.91 & 1.13 & O & 51 & 14 & He? & \nodata & \nodata \\
38 & WD0100-567 & 15.17948 & $-$56.77684 & 293 & 293 & 6 & 8 & O & 17.44 & 0.60 & O & 55 & 13 & H & \nodata & \nodata \\
\hline
A1 & WD2221-402 & 335.46833 & $-$40.19267 & 316 & $-$238 & 17 & 0 & U & 19.81 & 1.09 & O & 80 & 20 & He & \nodata & \nodata \\
A2 & WD2342-225 & 355.56890 & $-$22.45330 & 312 & 90 & 6 & 2 & U & 19.41 & 0.82 & O & 90 & 22 & He & \nodata & \nodata \\
A3 & WD0007-031 & 1.77802 & $-$3.11857 & 224 & $-$390 & 3 & 4 & U & 18.44 & 0.79 & O & 64 & 15 & H? & \nodata & \nodata \\
A4 & WD2236-168 & 339.06532 & $-$16.79833 & 318 & $-$60 & 3 & 6 & U & 18.48 & 0.75 & O & 69 & 17 & He & \nodata & \nodata \\
A5 & WD2234-408 & 338.72467 & $-$40.75506 & 287 & $-$249 & 17 & 14 & O & 17.72 & 0.47 & O & 76 & 18 & H & \nodata & \nodata \\
A6 & LHS 1044\tablenotemark{e} & 3.55330 & $-$13.18362 & $-$554 & $-$708 & 2 & 2 & U & 15.78 & 0.68 & O & 22 & 6 & H & \nodata & \nodata \\
A7 & J0424-4551\tablenotemark{f} & 65.99036 & $-$45.84513 & $-$100 & $-$532 & 15 & 1 & U & 16.85 & 0.73 & O & 34 & 8 & H & \nodata & \nodata \\
A8 & LHS 3917\tablenotemark{g} & 348.82831 & $-$2.16120 & 584 & 192 & 3 & 1 & U & 16.48 & 0.50 & O & 41 & 10 & He & \nodata & \nodata \\
A9 & LHS 4041 & 358.57837 & $-$36.56524 & 26 & $-$664 & 1 & 2 & U & 15.46 & $-$0.02 & O & 59 & 14 & H & $-$27 & 3 \\
A10 & JL 193 & 7.85905 & $-$44.63682 & 342 & 28 & 13 & 10 & U & 16.89 & 0.11 & O & 91 & 24 & He & \nodata & \nodata \\
A11 & LP 880-451\tablenotemark{h} & 1.78131 & $-$31.22642 & 336 & $-$124 & 3 & 6 & U & 16.47 & $-$0.12 & O & 108 & 26 & He & \nodata & \nodata \\
A12 & LHS 1076\tablenotemark{i} & 6.66973 & $-$55.41222 & $-$294 & $-$450 & 5 & 10 & U & 15.16 & 0.23 & O & 34 & 9 & H & \nodata & \nodata \\
A13 & WD0252-350\tablenotemark{j} & 43.65459 & $-$34.83158 & 44 & $-$328 & 2 & 3 & U & 15.79 & $-$0.05 & O & 71 & 17 & H & 86 & 2 \\
\enddata
\tablecomments{Objects below the line are the new cool WD candidates,
coming from OHDHS sample, but not listed in OHDHS paper. Coordinates
are given for epoch and equinox 2000. Radial velocities are corrected
for gravitational redshift and come from this work, except for A9 and A13
\citep{spy}.}  
\tablenotetext{a}{Astrometry source flag. U = USNO-B1.0, O = OHDHS.}  
\tablenotetext{b}{Photometry source flag. S = CCD photometry from this paper,
O = Calibrated from OHDHS (SuperCOSMOS) magnitudes.}
\tablenotetext{c}{Color-magnitude relation flag. H = Hydrogen CMR (DA WD), He
= Helium CMR, H? = Hydrogen CMR used, \ha insecure. He? = Helium CMR used, but could be a non-DA hydrogen WD. sp = Special.}
\tablenotetext{d}{As discussed in \ref{ssec:w}, \citet{lhs4033} are about to 
publish trigonometric parallax of this WD, showing it to be 30 pc distant and 
bringing down the redshift-corrected radial velocity to $76\kms$ (H. Harris, 
priv.\ comm.)}
\tablenotetext{e}{DA-type, $V=15.89$, $V-I=0.67$, $\pi_{\rm trig}=51.3\pm 3.8$ mas
\citep{berg01}.}
\tablenotetext{f}{DA9.5 \citep{scholz}.}
\tablenotetext{g}{DZ7.5, Villanova WD Catalog (online); $V=16.31$, $V-I=0.49$, $\pi_{\rm trig}=37.5\pm 5.9$ mas \citep{berg01}.}
\tablenotetext{h}{DB3, Villanova WD Catalog (online).}
\tablenotetext{i}{DA5, $V=15.14$, Villanova WD Catalog (online).}
\tablenotetext{j}{Name from \citet{spy}.}

\end{deluxetable} 

\end{document}